\documentclass[]{spie}  
\usepackage[]{graphicx}

\graphicspath{{./images_all/SPIE_2014/}}

\usepackage{amsmath}
\usepackage{amssymb}

\usepackage{cite}

\title{Ultra-low-energy straintronics using multiferroic composites} 

\author{Kuntal Roy
\skiplinehalf
School of Electrical and Computer Engineering\\Purdue University, West Lafayette, IN 47907, USA
}

\authorinfo{E-mail: royk@purdue.edu, Some works were performed prior to joining Purdue University.}

\begin{document} 
  \maketitle 

\begin{abstract}
The primary impediment to continued improvement of traditional charge-based electronic devices in accordance with Moore's law is the excessive energy dissipation that takes place in the devices during switching of bits. One very promising solution is to utilize strain-mediated multiferroic composites, i.e., a magnetostrictive nanomagnet strain-coupled to a piezoelectric layer, where the magnetization can be switched between its two stable states in sub-nanosecond delay while expending a minuscule amount of energy of $\sim$1 attojoule at room-temperature. Apart from devising digital memory and logic, these multiferroic devices can be also utilized for analog signal processing, e.g., voltage amplifier. First, we briefly review the recent advances on multiferroic straintronic devices and then we show here that in a magnetostrictive nanomagnet, it is possible to achieve the so-called Landauer limit (or the ultimate limit) of energy dissipation of amount $kT\,ln(2)$ compensating the entropy loss, thereby linking information and thermodynamics.
\end{abstract}



\section{INTRODUCTION}
\label{sec:intro}  

\begin{figure}
\begin{center}
\begin{tabular}{c}
\includegraphics{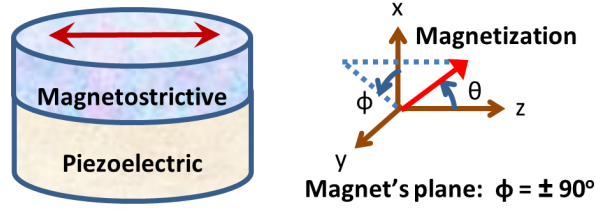}
\end{tabular}
\end{center}
\caption
{\label{fig:multiferroic_composite} 
\textbf{Schematic diagram of a strain-mediated multiferroic composite (piezoelectric-magnetostrictive heterostructure), and axis assignment.} The magnetostrictive nanomagnet is shaped like an elliptical cylinder and it has a single ferromagnetic domain. It's mutually anti-parallel magnetization states along the $z$-axis store the binary information 0 or 1. In standard spherical coordinate system, $\theta$ is the polar angle and $\phi$ is the azimuthal angle. According to the convention, we term the $z$-axis the easy axis, the y-axis the in-plane hard axis, and the $x$-axis the out-of-plane hard axis based on the chosen dimensions of the nanomagnet.}
\end{figure}

Electric field induced magnetization switching in multiferroic composites works according to the principles of converse magnetoelectric effect, i.e., when an electric field is applied across the structure, it induces a magnetic anisotropy, which can rotate the magnetization. This eliminates the need to utilize the cumbersome magnetic field or high spin-polarized current to rotate the magnetization.~\cite{roy13_spin,roy11_news} Hence, it can harness an energy-efficient binary switch replacing the traditional charge-based transistors for our future information processing systems.~\cite{roy13_spin,roy11_news} This turns out to be a very promising mechanism since a small voltage can generate a huge magnetic anisotropy, e.g., with suitable choice of materials, when a voltage of few millivolts is applied across a strain-mediated multiferroic composite device (see Fig.~\ref{fig:multiferroic_composite}), i.e., a magnetostrictive nanomagnet attached to a piezoelectric layer,~\cite{RefWorks:558,Refworks:164,Refworks:165,RefWorks:519} the piezoelectric layer is strained and the generated strain is transferred to the magnetostrictive layer. Then the induced stress anisotropy in the nanomagnet can switch the magnetization between its two stable states that store a binary information 0 or 1.~\cite{roy11_6,roy13_2,roy11_2}. This study has opened up a new field named straintronics~\cite{roy13_spin,roy13,roy14} and experimental efforts to realize such devices are considerably emerging.~\cite{RefWorks:559,RefWorks:806,RefWorks:609,RefWorks:790} Although the experimental efforts have demonstrated the induced stress anisotropy in the magnetostrictive nanomagnets, the direct experimental demonstration of switching delay (rather than ferromagnetic resonance experiments~\cite{RefWorks:806}) and utilizing low-thickness piezoelectric layers [e.g., $<$ 100 nm of lead magnesium niobate-lead titanate (PMN-PT)~\cite{RefWorks:820}] for ultra-low energy dissipation are still under investigation.


While utilizing strain-mediated multiferroic composites for \emph{digital} binary switching of the magnetization has been investigated, the \emph{analog} signal processing capability using inherently digital nanomagnets has not been thought of. It is true that computing and signal processing tasks are mostly shifted to digital domain, however, sometimes analog signal processing is fundamentally necessary, e.g., processing of natural signals.~\cite{razav01} When a transmitted signal is received at the receiver end, the signal is weak in magnitude and also noisy due to attenuation and noise in the environment. Hence, the signal needs to be amplified and filtered. Only after converting the analog signal to digital domain, we can process the signal digitally. Therefore, for different purposes, we do have the requirement of analog signal processing. In a very recent development, it is shown that we can utilize these multiferroic straintronic devices for ultra-low-energy analog signal processing too.~\cite{unpub2x} A voltage applied across a multiferroic composite induces stress on the magnetostrictive layer and this modulates the potential landscape of the nanomagnet. To be able to harness the analog nature, we need to rotate the magnetization \emph{continuously} rather than having an abrupt switching as required by digital computation. The \emph{continuous} rotation of magnetization is conceived by considering that magnetization is \emph{not} exactly stable at a position rather it is fluctuating around a mean value due to thermal agitations. The applied voltage modulates the potential landscape of the magnetostrictive nanomagnet and the mean value of magnetization can be changed \emph{continuously}. Using tunneling magnetoresistance (TMR) measurement,~\cite{RefWorks:577,RefWorks:33,RefWorks:581} this continuous mean value can be compared with respect to a fixed nanomagnet and a continuous output voltage while varying the input voltage can be produced. Stochastic Landau-Lifshitz-Gilbert (LLG) equation of magnetization dynamics at room-temperature~\cite{RefWorks:162,RefWorks:161,RefWorks:186} is solved to demonstrate this concept of analog signal processing functionality.~\cite{unpub2x}

\subsection{Theory and model} 

It is a common practice to model nanomagnets as shape-anisotropic elliptical cylinders (see Fig.~\ref{fig:multiferroic_composite}) to produce an energy barrier between its two stable states anti-parallel to each other along the elongated axis ($z$-axis). When stress is generated on the magnetostrictive nanomagnet, the total potential energy of the stressed polycrystalline (we neglect the magnetocrystalline anisotropy) single-domain nanomagnet per unit volume is the sum of the shape anisotropy energy and the stress anisotropy energy:~\cite{roy13_spin}
\begin{align}
E_{total}(\theta,\phi,t) &= E_{shape}(\theta,\phi) + E_{stress}(\theta,t) \nonumber\\
												 &= B_{shape}(\phi) \, sin^2\theta  - B_{stress}(t)\,cos^2\theta,
\label{eq:E_total_ss}
\end{align}
where
\begin{align}
B_{shape}(\phi) &= (1/2) \, M \, [H_k  + H_d\, cos^2\phi], \label{eq:B_shape}\\
B_{stress}(t) &= (3/2) \, \lambda_s \sigma(t), \label{eq:B_stress}
\end{align}
\noindent
$M=\mu_0 M_s$, $\mu_0$ is permeability of free space, $M_s$ is the saturation magnetization, $H_k=(N_{dy}-N_{dz})\,M_s$ is the Stoner-Wohlfarth switching field,~\cite{RefWorks:557} $H_d=(N_{dx}-N_{dy})\,M_s$ is the out-of-plane demagnetization field,~\cite{RefWorks:157} $N_{dm}$ is the m$^{th}$ ($m=x,y,z$) component of the demagnetization factor~\cite{RefWorks:402} ($N_{dx} \gg N_{dy} > N_{dz}$), $(3/2) \lambda_s$ is the magnetostriction coefficient of the magnetostrictive nanomagnet,~\cite{RefWorks:157} $\sigma=Y\epsilon$ is the stress on the nanomagnet, $Y$ is the Young's modulus, $\epsilon$ is the strain that generates the stress.

Materials having positive $\lambda_s$ (e.g., Terfenol-D, Galfenol) requires a compressive (negative by convention) stress to favor the alignment of magnetization along the minor axis ($y$-axis in Fig.~\ref{fig:multiferroic_composite}), while the materials having negative $\lambda_s$ (e.g., Fe, Co, Ni) requires a tensile (positive by convention) stress for the same. The gradient of potential profile due to shape and stress anisotropy exerts an effective field on the magnetization and this generates a torque $\mathbf{T_E}$,~\cite{roy13_spin} while the torque due to thermal fluctuations $\mathbf{T_{TH}}$ is treated via a random magnetic field.~\cite{roy13_spin,RefWorks:186}


In the macrospin approximation, the magnetization \textbf{M} of the nanomagnet is constant in magnitude but it varies with direction, so that we can represent it by a vector of unit norm $\mathbf{n_m} =\mathbf{M}/|\mathbf{M}|$. The magnetization dynamics under the action of these two torques $\mathbf{T_E}$ and $\mathbf{T_{TH}}$ is described by the stochastic Landau-Lifshitz-Gilbert (LLG) equation~\cite{RefWorks:162,RefWorks:161,RefWorks:186} as follows:
\begin{equation}
\label{eq:LLG}
\cfrac{d\mathbf{n_m}}{dt}-\alpha\left(\mathbf{n_m} \times \cfrac{d\mathbf{n_m}}{dt}\right) = - \cfrac{|\gamma|}{M}\,\left\lbrack \mathbf{T_E} + \mathbf{T_{TH}}\right\rbrack,
\end{equation}
where $\alpha$ is the phenomenological Gilbert damping parameter~\cite{RefWorks:161} through which magnetization relaxes to the minimum energy position, and $\gamma$ is the gyromagnetic ratio of electrons. Note that $\mathbf{M}$, $\mathbf{T_E}$, and $\mathbf{T_{TH}}$ are all proportional to the nanomagnet's volume. Solving the above equation, we can track the trajectory of magnetization over time.

For the magnetostrictive layer, we need to choose a material that maximizes the product $(3/2)\lambda_s\, Y$. Terfenol-D (TbDyFe), which has 30 times higher magnetostriction coefficient in magnitude than the common ferromagnetic materials (e.g., Fe, Co, Ni), has the highest $(3/2)\lambda_s\, Y$.~\cite{roy11_2} If it needs to avoid the rare-earth materials (e.g., Tb and Dy in Terfenol-D), we can also utilize Galfenol (FeGa),~\cite{RefWorks:167,RefWorks:801} which has 6 times less $(3/2)\lambda_s$, but twice high $Y$ than that of Terfenol-D.~\cite{roy11_2}


For the piezoelectric layer, we may use lead-zirconate-titanate (PZT), but using lead magnesium niobate-lead titanate (PMN-PT) is preferable since it has high piezoelectric coefficient and it can generate anisotropic strain, which allows us to work with lower voltage for a required strain, thereby reducing the energy dissipation.  PMN-PT layer has a dielectric constant of 1000, $d_{31}$=--3000 pm/V, and $d_{32}$=1000 pm/V.~\cite{RefWorks:790} With the piezoelectric layer's thickness $t_{piezo}$=24 nm,~\cite{roy11_6} $V=1.9$ mVs (2.9 mVs) of voltages would generate 20 MPa (30 MPa) compressive stress [$\sigma=Y\,d_{eff}\,(V/t_{piezo})$, where $d_{eff}=(d_{31}-d_{32})/(1+\nu)$] in the magnetostrictive Terfenol-D layer, which has $Y=80$ GPa,~\cite{roy11_6} and Poisson's ratio $\nu=0.3$.~\cite{RefWorks:821}. Modeling the piezoelectric layer as a parallel plate capacitor ($\sim$100 nm lateral dimensions), the capacitance C=2.6 fF and thus $CV^2$ energy dissipation turns out to be $<$ 0.1 aJ. This is the basis of ultra-low-energy computing using these multiferroic devices.~\cite{roy11_news,roy13_spin,roy13,roy14}

We do consider the distribution (rather than a fixed value) of initial orientation of magnetization due to thermal fluctuations, which has crucial consequence on device performance.~\cite{roy13_2,roy11_6,roy14} Note that we have assumed here \emph{uniform} magnetization and \emph{uniform} strain in the magnetostrictive nanomagnet. This is quite valid in small length scales and such assumptions allow us to get critical insights on the device operation. A detailed space-dependent solution may be sought for \emph{quantitative} purposes, however, it would require an enormous time to execute. Hence, we stick to the aforesaid assumptions.

\begin{figure}
\begin{center}
\begin{tabular}{c}
\includegraphics[width=0.95\textwidth]{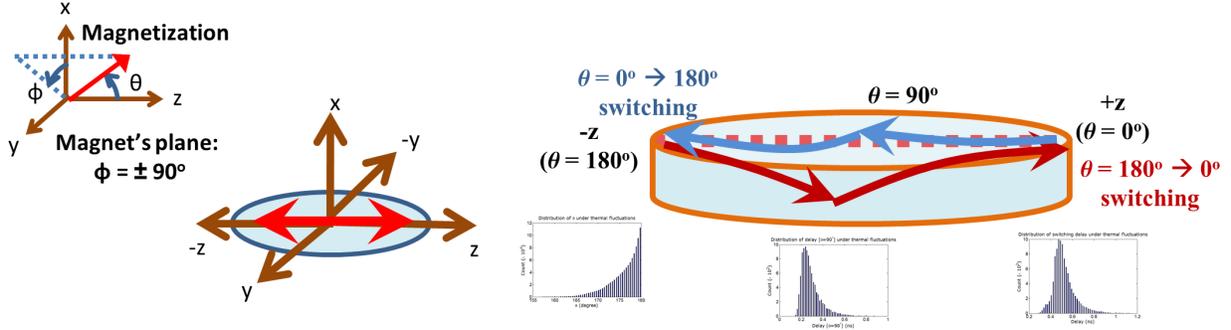}
\end{tabular}
\end{center}
\caption
{\label{fig:digital_memory} 
\textbf{Digital memory using multiferroic straintronic devices.} 
Switching of magnetization between two anti-parallel states ($\pm z$-axis). Stress rotates magnetization out of magnet's plane and when magnetization reaches the hard plane ($\theta=90^\circ$), intrinsic dynamics dictates that magnetization lifts out-of-plane in a certain direction so that a complete $180^\circ$ switching of magnetization is possible.~\cite{roy13_2}. While switching along the --$y$-axis rather than +$y$-axis (shown by arrows), the directions of the out-of-plane excursions would be exactly opposite.~\cite{roy13_2} Three distributions are shown: the one at $\theta=180^\circ$ depicts the initial distribution when no stress is active, the other two distributions around $\theta=90^\circ$ and $\theta=0^\circ$ correspond to 60 ps ramp period and 15 MPa stress.~\cite{roy13_2}}
\end{figure}

\subsection{Digital memory}

Here we briefly review the binary switching of magnetization in the magnetostrictive nanomagnets for devising digital memory devices.~\cite{roy13_2} The usual perception is that stress can rotate magnetization only by $90^\circ$ from $\pm z$-axis to $\pm y$-axis (see Fig.~\ref{fig:digital_memory}). However, the torque due to stress acts along the out of magnet's plane ($\mathbf{\hat{e}_\phi}$ direction) and therefore magnetization lifts out-of-plane (although the out-of-plane demagnetization field due to small thickness of nanomagnet tries to keep the magnetization in-plane).
\begin{equation}
\mathbf{T_{E,stress}} = - \mathbf{\hat{e}_r} \times \nabla\,E_{stress}  = - (3/2) \, \lambda_s \sigma sin(2\theta)\, \mathbf{\hat{e}_\phi}.
\label{eq:T_stress}
\end{equation}
\noindent
This out-of-plane excursion of magnetization generates an \emph{intrinsic} asymmetry, which can completely switch the magnetization by $180^\circ$.~\cite{roy13_2} Full $180^\circ$ switching is desirable since it facilitates having the full tunneling magnetoresistance (TMR) while electrically reading the magnetization state using a magnetic tunnel junction (MTJ). As depicted in the Fig.~\ref{fig:digital_memory}, magnetization's initial distribution due to thermal fluctuations is quite wide.~\cite{roy13_2} Also, thermal fluctuations create a distribution for the time required to reach at the hard plane ($\theta=90^\circ$, $y$-$z$ plane). Therefore, to tackle this wide distribution, it requires a sensing circuitry to detect when magnetization reaches around $\theta=90^\circ$ and withdraw/reverse stress subsequently to be able to switch $180^\circ$ successfully.~\cite{roy13_2} The simulation results show that fast ramp rates and high stresses are conducive to successful switching.~\cite{roy13_2,roy11_6} It should be noted that such switching leads to a \emph{toggle} memory unless we have a mechanism to maintain the direction of switching.~\cite{roy14_2,RefWorks:649} According to Ref.~\citenum{roy14_2}, the interface coupling between the polarization and magnetization can maintain the direction of switching. Also the strong coupling facilitates error-resilient switching without the need of having a sensing circuitry and it can lead to lowering the lateral dimensions of the nanomagnet to $\sim$10 nm. The $90^\circ$ switching mechanisms can be also utilized to direct switching in a particular direction, however, it gives us lower TMR. 

On reading the magnetization state of the magnetostrictive nanomagnet, a material issue crops up since magnetostrictive materials (Terfenol-D, Galfenol) cannot be in general used as the free layer of an MTJ. Usually, CoFeB is used as the free layer alongwith MgO spacer that gives us high TMR,~\cite{RefWorks:33} while using half-metals can lead to even higher TMR.~\cite{RefWorks:581} This issue can be simply solved by magnetically couple the magnetostrictive layer and the free layer, e.g., exchange coupling or introducing an insulator between the layers and exploiting the dipole coupling in between to rotate them concomitantly.~\cite{unpub}

%

\begin{figure}
\begin{center}
\begin{tabular}{c}
\includegraphics{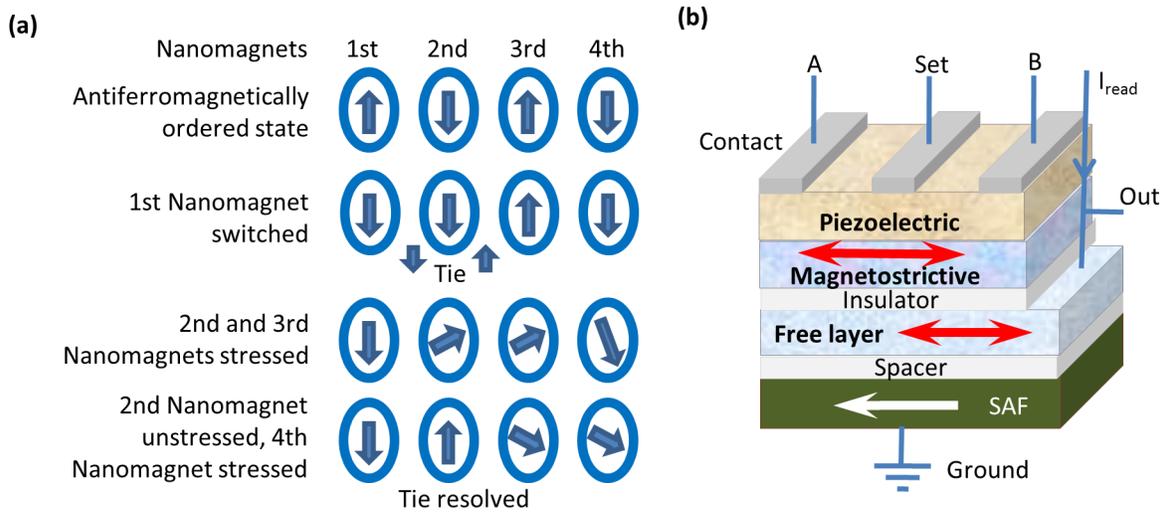}
\end{tabular}
\end{center}
\caption
{\label{fig:digital_logic} 
\textbf{Digital logic design using multiferroic straintronic devices.}
(a) Information propagation through a horizontal chain of straintronic devices \emph{uni-directionally}. Since the dipole coupling between the neighboring nanomagnets is \emph{bidirectional}, we need to impose the unidirectionality in time (using a 3-phase clocking scheme) to subsequently clock the nanomagnets in the chain.~\cite{roy14} With this scheme, logic gates like NAND and NOR, and complex circuits can be devised. (Reprinted with permission from Ref.~\citenum{roy14}. Copyright 2014, AIP Publishing LLC.) (b) Logic design using read-write units. The read unit is an MTJ, which reads the free layer's magnetization direction with respect to the fixed layer. The magnetostrictive layer and the free layer are dipole coupled (anti-parallel orientation), hence the magnetostrictive nanomagnet's direction can be read. Such read-write units can be concatenated to devise complex circuits. Two inputs can be incorporated to design single-element straintronic universal logic gates (e.g., NAND, NOR).~\cite{roy13} By applying voltages at the terminals A and B, the magnetization of the magnetostrictive nanomagnet can be switched and depending on the design of the nanomagnets, the respective logic operation can be performed.~\cite{roy13}}
\end{figure} 

\subsection{Digital logic}

Apart from devising memory bits, these straintronic multiferroic devices have been proposed for logic design purposes too.~\cite{roy13_spin,roy13,roy14} Figure~\ref{fig:digital_logic}a depicts Bennett clocking mechanism~\cite{RefWorks:144} of information propagation through a horizontal chain of multiferroic straintronic devices.~\cite{roy13_spin} There is no wired connection between the nanomagnets and the information transfer happens due to dipole coupling between the nanomagnets. While considering room-temperature operation, the intrinsic dynamics due to out-of-plane excursion of magnetization causes switching failures~\cite{roy14} and it can remedied by using the intrinsic asymmetry, i.e., releasing/reversing stress dynamically when magnetization reaches the hard plane ($\theta=180^\circ$)~\cite{roy14} or exploiting the asymmetry due to interface coupling.~\cite{roy14_2}

An intriguing mechanism to build logic is to use the concept of read-write units.~\cite{roy13} The written bit in the magnetostrictive nanomagnet is read by an MTJ (read-unit) and the output can be fed to next write units (multiferroic composites), i.e., individual read-write units can be concatenated since they have voltage gain and input-output isolation.~\cite{roy13} Also, by utilizing two inputs as shown in the Fig.~\ref{fig:digital_logic}b, 2-input universal logic gates (NAND and NOR) can be designed.~\cite{roy13} Higher input logic gates and majority logic gates can be devised according to the same concept.~\cite{roy13} This design methodology can overwhelmingly simplify the design of a large scale circuit and portend a highly dense yet an ultra-low-energy computing paradigm. 

\begin{figure}
\begin{center}
\begin{tabular}{c}
\includegraphics[width=0.9\textwidth]{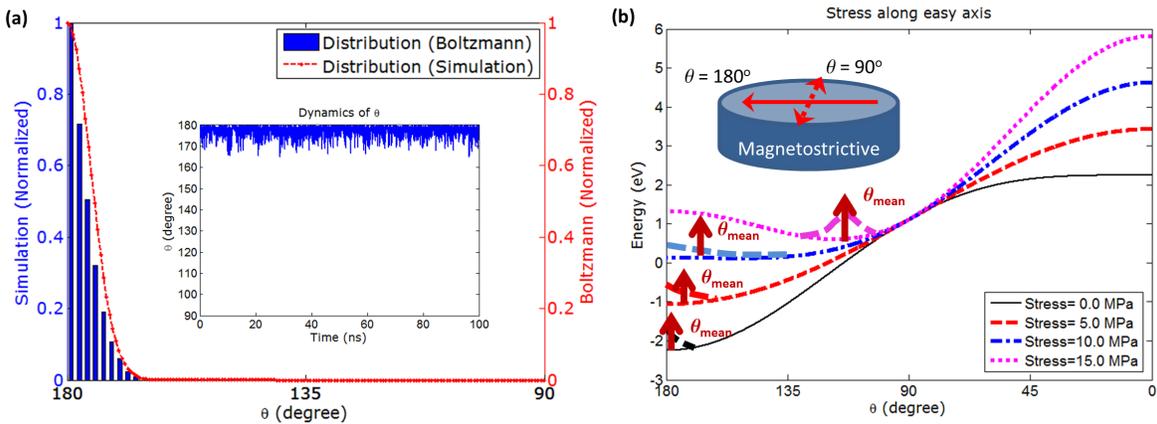}
\end{tabular}
\end{center}
\caption
{\label{fig:analog_distribution} 
\textbf{Continuous rotation of magnetization in magnetostrictive nanomagnets considering thermal fluctuations.}
(a) Magnetization is fluctuating due to room-temperature (300 K) thermal agitations around one easy axis $\theta=180^\circ$. The distribution is achieved by solving stochastic LLG at 300 K. The distribution turns out to be a Boltzmann distribution as expected. (b) The potential landscape of the magnetostrictive nanomagnet is modulated with stress. As the energy barrier decreases, the mean orientation of the magnetization changes \emph{continuously} rather than abruptly. When sufficient stress is exerted, magnetization comes toward the hard-axis ($\theta=90^\circ$). Note that the potential landscape of the magnetostrictive nanomagnet is made monostable (while no stress is active) by having a magnetic coupling from a neighboring nanomagnet to avoid abrupt switching. }
\end{figure}

\begin{figure}
\begin{center}
\begin{tabular}{c}
\includegraphics{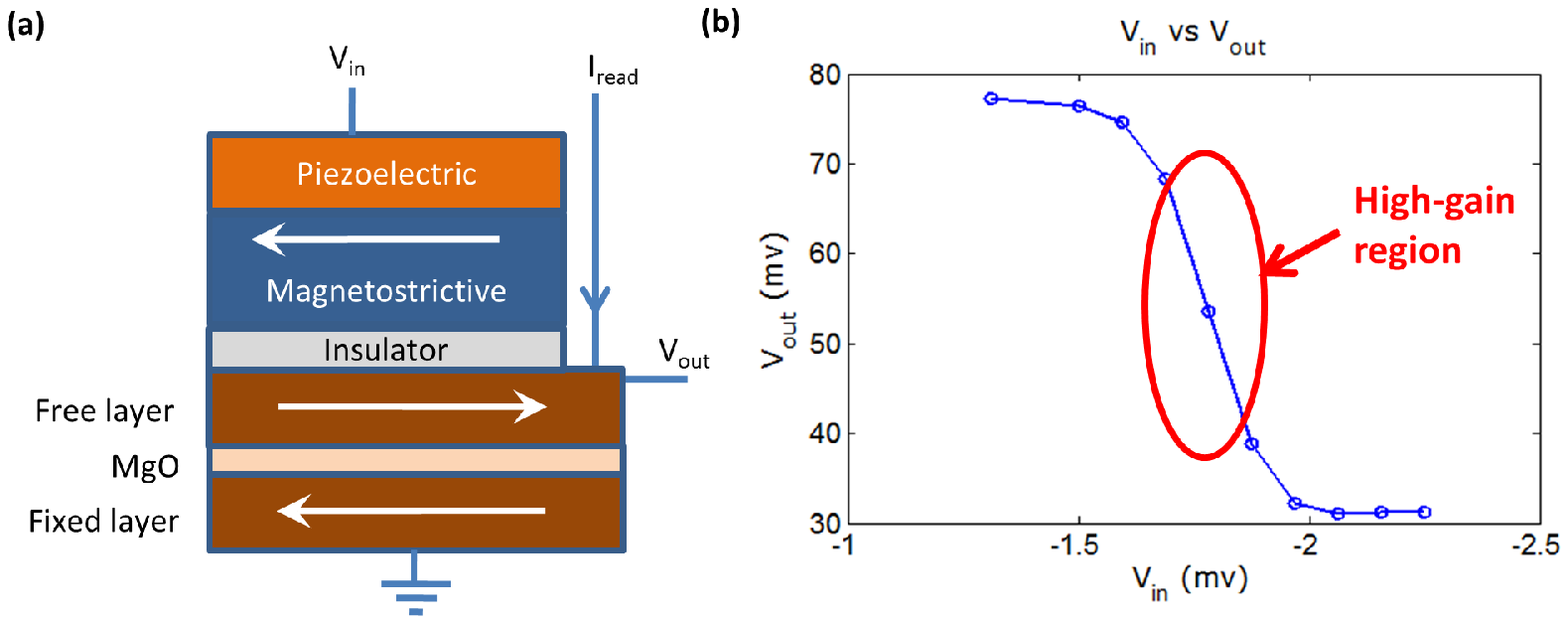}
\end{tabular}
\end{center}
\caption
{\label{fig:analog_gain} 
\textbf{Design of a voltage amplifier using multiferroic straintronic devices.}
(a) The device structure to harness the analog nature consists of a multiferroic composite attached to an MTJ separated by an insulator. The magnetostrictive nanomagnet is magnetically coupled to the free layer via dipole coupling. $I_{read}$ is a constant read current and the output voltage $V_{out}$ varies continuously with $V_{in}$. (b) The relation between the input voltage $V_{in}$ and the output voltage $V_{out}$ for a Terfenol-D/PMN-PT multiferroic composite. This is an inverter like characteristics with a high-gain transition region. If we bias the input voltage in this high-gain region, we would get an amplified output voltage.}
\end{figure} 

\subsection{Analog signal processing}

While a bistable double-well separated by an energy barrier is desired for digital computing, analog signal processing requires a \emph{continuous} rotation of magnetization. The potential landscape of a shape-anisotropic single-domain nanomagnet is a symmetric double-well and hence it needs to stop the abrupt switching between the two stable states for \emph{analog} signal processing. We can make the potential landscape \emph{monostable} by utilizing magnetic coupling from a neighboring nanomagnet. As depicted in Fig.~\ref{fig:analog_distribution}, we choose the potential well at $\theta=180^\circ$. Figure~\ref{fig:analog_distribution}a shows magnetization's fluctuations due to room-temperature (300 K) thermal agitation by solving stochastic LLG and it fits quite well with the Boltzmann distribution as expected. Figure~\ref{fig:analog_distribution}b shows that with the application of stress, magnetization's mean orientation shifts \emph{continuously} and when the stress is sufficient enough, magnetization comes close to $\theta=90^\circ$. An amplifier with a voltage gain can be designed utilizing this methodology as depicted in the Fig.~\ref{fig:analog_gain}.



We can directly put an AC signal at the input terminal of the device (see Fig.~\ref{fig:analog_gain}a), solve stochastic LLG, and the output voltage can be extracted from the TMR measurement of the MTJ. For an AC signal with 1 GHz frequency, a voltage gain of 50 is achieved, while dissipating a miniscule amount of energy of $\sim$0.1 attojoule/cycle at room-temperature.~\cite{unpub2x} A rectifier characteristics can be obtained from these straintronic devices too since with the application of a voltage of opposite polarity (that generates stress of opposite polarity too, see Fig.~\ref{fig:analog_distribution}b), magnetization becomes more confined in the $\theta=180^\circ$ well. The degree of rectification capability increases with the TMR of the MTJ.~\cite{unpub2x}

\section{LANDAUER LIMIT OF ENERGY DISSIPATION} 

Computers are physical systems and therefore a computation or processing of information is subjected to physical principle, e.g., the second law of thermodynamics. According to Landauer's principle,~\cite{RefWorks:148} in a classical system with two degenerate ground states, a minimum amount of energy proportional to temperature $kT ln(2)$ ($\sim3 \times 10^{-21}$ joule at T=300 K, $k$ is the Boltzmann constant, and T is temperature) must be dissipated to \emph{erase} a bit of information compensating the entropy loss, thereby linking the information and thermodynamics. Maxwell introduced his controversial \emph{demon}~\cite{maxwell,RefWorks:615,RefWorks:619,RefWorks:642} over a century ago elucidating the relationship between information and entropy, however, the famous demon was later exorcized by Bennett.~\cite{RefWorks:614,RefWorks:627,RefWorks:528,RefWorks:527,RefWorks:645,RefWorks:613,RefWorks:522,RefWorks:144,RefWorks:644} Note that it is possible to have  \emph{stochastic} violation of Landauer's principle in small systems, since they are prone to thermal fluctuations.~\cite{RefWorks:629,RefWorks:637,RefWorks:626,RefWorks:623,RefWorks:624,RefWorks:643} Stochastic violations of second law have been experimentally observed,~\cite{RefWorks:621,RefWorks:622} however, the second law is still safeguarded \emph{on average}.~\cite{RefWorks:620} 

Recently, Landauer limit of energy dissipation is experimentally demonstrated for a colloidal particle with \emph{linear} motion.~\cite{RefWorks:617} However, the Landauer's limit is not experimented for a \emph{rotational} body like the magnetization in a nanomagnet.~\cite{RefWorks:557,RefWorks:157,RefWorks:490} With the advent of experimental apparatus, \emph{single-domain} nanomagnets~\cite{RefWorks:133,RefWorks:402,RefWorks:426} having two stable states separated by an energy barrier can become the staple of future information processing systems (since the excessive energy dissipation during switching of bits has been the bottleneck behind utilizing traditional charge-based transistor electronics further~\cite{kilby,moore65,RefWorks:211}). Particularly, magnetostrictive nanomagnets in multiferroic heterostructures have profound potential to act as the basic building block in ultra-low-energy computing systems.~\cite{roy13_spin,roy13} Here, we show that Landauer limit of energy dissipation is achievable in magnetostrictive nanomagnets. It is pointed out that magnetization may deflect out of magnet's plane during its dynamical motion and even a very small out-of-plane excursion plays a crucial role in shaping the magnetization dynamics. Therefore, it is imperative to consider the complete three-dimensional potential landscape and full three-dimensional motion of the magnetization, rather than \emph{assuming} an overdamped particle with \emph{linear} motion.~\cite{RefWorks:623,RefWorks:624} We have solved stochastic Landau-Lifshitz-Gilbert (LLG) equation~\cite{RefWorks:162,RefWorks:161,RefWorks:186} in the presence of thermal fluctuations to determine the energy dissipation during the erasure of a bit of information and show that stochastic violation of the Landauer bound is possible, nonetheless the bound is respected \emph{on average}.

\begin{figure}
\begin{center}
\begin{tabular}{c}
\includegraphics{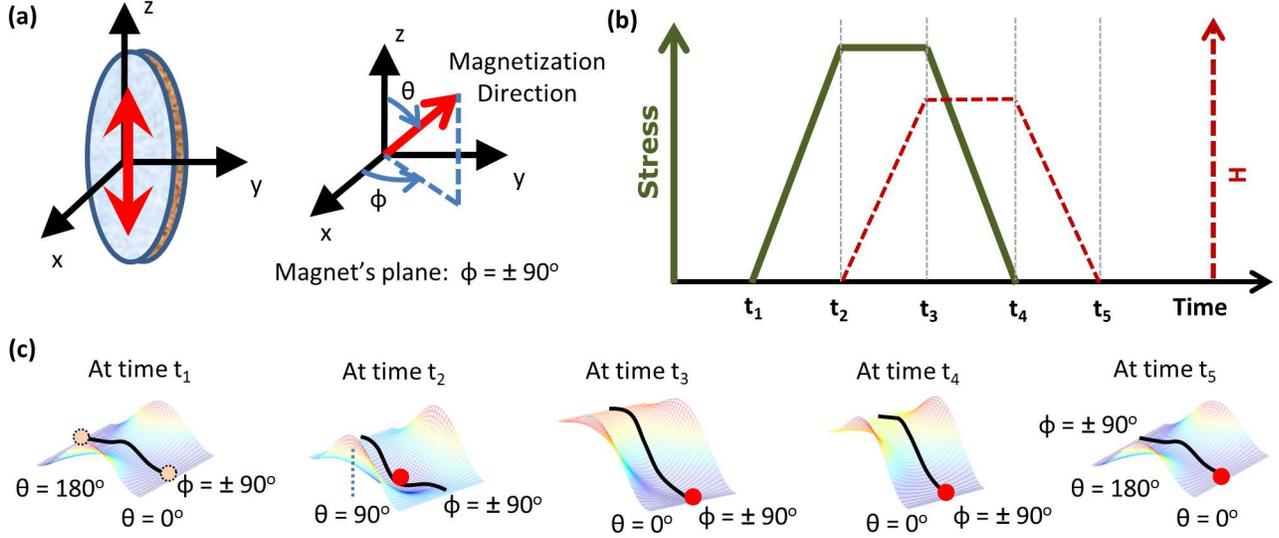}
\end{tabular}
\end{center}
\caption
{\label{fig:magnetization_potential} 
\textbf{Deterministic switching of magnetization to a final stable state in a magnetostrictive nanomagnet.}
(a) A shape-anisotropic single-domain magnetostrictive nanomagnet having two stable states along the elongated direction and axis assignment. (b) The time-cycle of uniaxial stress (along $z$-direction) and an asymmetry making field, \textit{H} (along $\theta=0^\circ$ direction) acting on the magnetostrictive nanomagnet. (c) The complete three-dimensional potential landscapes of the nanomagnet at different instants of time. The solid lines on the landscapes correspond to the in-plane ($y$-$z$ plane, $\phi=\pm90^\circ$) potential profiles. Although the in-plane potential landscapes are the minimum energy positions, the out-of-plane potential landscapes do play an important role in shaping the magnetization dynamics as discussed in the text. At time $t_1$, magnetization can reside in either of the wells ($\theta=0^\circ$ or $\theta=180^\circ$) having 50\% probability each, however, at time $t_2$, it reaches at $\theta=90^\circ$, i.e., the bit is \emph{erased}. At time $t_3$ and onwards, magnetization reaches at $\theta \simeq 0^\circ$ \emph{deterministically} due to the asymmetry making field.}
\end{figure} 

\subsection{Model}

Figure~\ref{fig:magnetization_potential}a shows a single-domain magnetostrictive nanomagnet shaped like an elliptical cylinder with its elliptical cross-section lying on the $y$-$z$ plane; the major axis and minor axis are aligned along the $z$- and $y$-direction, respectively. In standard spherical coordinate system, $\theta$ is the polar angle and $\phi$ is the azimuthal angle. The magnet's plane is $y$-$z$ plane ($\phi = \pm90^{\circ}$). Any deviation of magnetization from magnet's plane is termed as out-of-plane excursion. The dimensions of the major axis, the minor axis, and the thickness are $a$, $b$, and $l$, respectively. So the magnet's volume is $\Omega=(\pi/4)abl$. Due to shape anisotropy, the two anti-parallel degenerate states $\theta=0^\circ$ and $\theta=180^\circ$ along the $z$-direction (easy axis) can store a binary bit of information $0$ or $1$. The $y$-axis and $x$-axis are the in-plane and the out-of-plane hard axis, respectively. Since $l \ll b$, the out-of-plane hard axis is much harder than the in-plane hard axis.

Figure~\ref{fig:magnetization_potential}b shows the time-cycle of uniaxial stress (along $z$-direction) and an asymmetry making field $H$ (along $\theta=0^\circ$ direction) acting on the magnetostrictive nanomagnet. The adiabatic pulses should be slow enough that magnetization follows its potential landscape and the quasistatic assumption is valid. We can write the total energy of the nanomagnet per unit volume as the sum of three energies -- the shape anisotropy energy, anisotropies induced due to applied stress and asymmetry-making magnetic field -- as follows:~\cite{RefWorks:157}
\begin{equation}
E_{total,asymm}(\theta,\phi,t) = B_{shape}(\phi) \, sin^2\theta  - B_{stress}(t)\,cos^2\theta - B_{asymm}(t) \, cos\,\theta,
\label{eq:E_total_asymm}
\end{equation}
where $B_{shape}(\phi)$ and $B_{stress}(t)$ are given by the Equations~\eqref{eq:B_shape} and~\eqref{eq:B_stress}, respectively, $B_{asymm}(t) = M H(t)$, and $H(t)$ is the asymmetry making magnetic field at time $t$. Note that it is possible to harness the asymmtery making field by intrinsic interface coupling between polarization and magnetization in multiferroic heterostructures.~\cite{RefWorks:649,roy14_2}


When a sufficiently high stress (of right polarity making the product $\lambda_s \sigma$ negative) is generated on the magnetostrictive nanomagnet and the in-plane shape-anisotropic energy barrier is overcome, magnetization will rotate toward the in-plane hard axis ($\theta=90^\circ$, $\phi=\pm90^\circ$). This is depicted in Fig.~\ref{fig:magnetization_potential}c. Magnetization initially can situate at any of the two \emph{degenerate} stable states ($\theta=0^\circ$, state $0$ and $\theta=180^\circ$, state $1$ with probabilities, $p_0 = p_1 = 0.5$). The initial entropy of the system at time $t_1$ is $-k\sum_n p_n\,ln\,p_n = k\, ln(2)$. However, when stress is applied between times $t_1$ and $t_2$, the barrier separating the two stable states gets removed and the potential landscape becomes monostable in $\theta$-space (at $\theta=90^\circ$) with entropy zero. A reduction of entropy $k\, ln(2)$ must be dissipated as heat during this process, according to the Landauer's principle. Note that the barrier is removed before applying any asymmetry-making field $H$. The rationale behind is to make the erasure process independent of barrier height, which determines the hold failure probability and also to resist the thermal fluctuations by making the monostable well deep enough.

Between times $t_2$ and $t_3$, an asymmetry-making field $H$ is applied to deterministically rotate the magnetization to state $\theta=0^\circ$. This is depicted in Fig.~\ref{fig:magnetization_potential}c. The asymmetry-making field tilts the potential landscape and the degree of this tilt should be sufficient enough to dissuade thermal fluctuations. Thereafter, both the stress and $H$ are removed to complete the switching process. Note that the entropy of the system at time $t_2$ and onwards are zero. Hence there is no bound on minimum energy dissipation when magnetization traverses from time $t_2$ to $t_3$ as the dissipation can be made arbitrarily small. Assuming  $(3/2)\lambda_s \sigma_{max} = - M H_k$ and $\phi=\pm90^\circ$, from Equation~\eqref{eq:E_total_asymm}, the total energy becomes $E_{total,asymm}(\theta,t)=-(1/2)M H_k\,sin^2\theta-M H(t)\,cos\theta$. We notice that as $H$ goes from $0$ to $H_k$, the minimum value of $\theta$ goes from $\theta=90^\circ$ to $\theta=0^\circ$ \emph{continuously} following the expression $\theta_{min}(t)=cos^{-1}(H(t)/H_k)$. 

With adiabatic pulses between times $t_1$ and $t_3$, the motion of magnetization is \emph{smooth} and magnetization follows the minimum potential energy landscape. Therefore there is no lower bound of energy dissipation \emph{in the absence of thermal fluctuations}. However, the Landauer's principle remains intact since this is $T=0\,K$ case and thus the energy dissipation proportional to temperature is also zero. If we incorporate random thermal fluctuations at finite temperatures, magnetization will get deflected uphill in the potential landscape even with pulses of very slow ramp and will incur energy dissipation, which is subjected to Landauer bound.

It needs mention here that particularly in the presence of thermal fluctuations, magnetization may temporarily traverse on higher potential not only in-plane of the nanomagnet but also \emph{out-of-plane} (i.e., when $\phi \neq \pm90^\circ$) and dissipate energy when it comes back to the lower potential. The simulation of magnetization dynamics results that a less than one degree of  deflection in the out-of-plane direction can have an immense consequence. The key reason behind is that the out-of-plane demagnetization field $H_d$ is about a couple of orders of magnitude higher than $H_k$. When the magnetization gets deflected out-of-plane due to torque exerted on it, an additional torque of comparatively very high magnitude due to $H_d$ comes into play, which makes the dynamics \emph{fast}. It is true that this out-of-plane excursion causes power dissipation but switching also becomes fast, so that the net energy dissipation may be smaller compared to the case when it is \emph{assumed} that magnetization is confined to the magnet's plane. Since the Landauer limit of energy dissipation is very small, we should particularly take into account this significant effect due to the out-of-plane excursion of magnetization, considering realistic parameters for magnetization, e.g., magnetization damping $\alpha$.

The torque $\mathbf{T_E}$ acting on the magnetization is derived from the gradient of potential landscape. Additionally, there is a random thermal field to incorporate thermal fluctuations.~\cite{RefWorks:186} We solve the stochastic Landau-Lifshitz-Gilbert equation~\cite{RefWorks:162,RefWorks:161,RefWorks:186} of magnetization dynamics [Equation~\eqref{eq:LLG}] and calculate the energy dissipation during the erasure of a bit of information. The energy dissipated in the nanomagnet due to Gilbert damping can be expressed as  $E_d = \int_0^{\tau}P_d(t) dt$, where $\tau$ is the time taken during the erasure cycle [i.e., $t_5-t_1$ in Fig.~\ref{fig:magnetization_potential}b], and $P_d(t)$ is the power dissipated at time $t$ per unit volume given by~\cite{roy13_spin}
\begin{equation}
P_d(t) = \frac{\alpha \, |\gamma|}{(1+\alpha^2) M} \, |\mathbf{T_E} (\theta(t), \phi(t), t)|^2.
\label{eq:power_dissipation}
\end{equation}
\noindent
Thermal field with mean zero does not cause any net energy dissipation but it causes variability in the energy dissipation by scuttling the trajectory of magnetization.

%


\begin{figure}
\begin{center}
\begin{tabular}{c}
\includegraphics[width=0.6\textwidth]{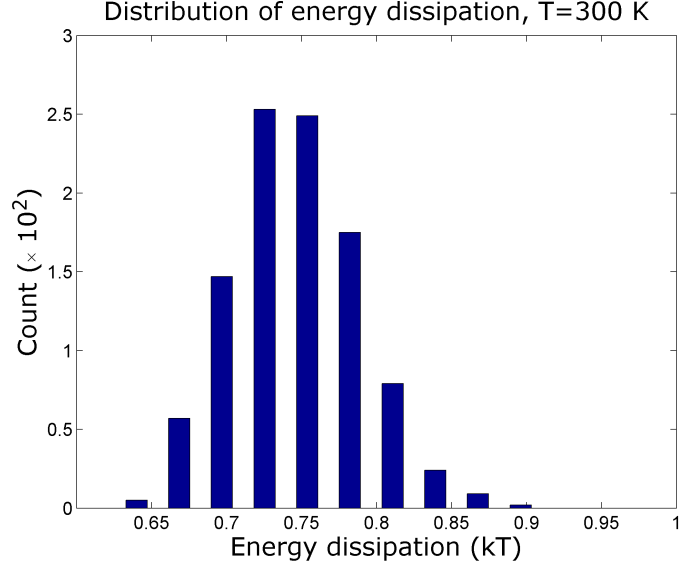}
\end{tabular}
\end{center}
\caption
{\label{fig:distribution_energy_300K} 
\textbf{Simulation results for Landauer limit of energy dissipation in a magnetostrictive nanomagnet.}
Distribution of energy dissipation at room-temperature (300 K) during the time interval $t_2 - t_1$. It is important to note that for a few cases, the Landauer bound ($kT\,ln(2)=0.6932\,kT$) is violated but the mean of the energy dissipation ($0.74\,kT$) exceeds the Landauer bound safeguarding the Landauer's principle and the second law of thermodynamics. A moderately large (1000) number of simulations in the presence thermal fluctuations have been performed to generate this distribution.}
\end{figure}

\subsection{Results}

The magnetostrictive nanomagnet is considered to be made of polycrystalline Galfenol (FeGa), which has the following material properties -- magnetostrictive coefficient ($(3/2)\lambda_s$): +150$\times$10$^{-6}$, saturation magnetization ($M_s$):  8$\times$10$^5$ A/m, Young's modulus (Y): 140 GPa, and  Gilbert damping parameter ($\alpha$): 0.025.~\cite{RefWorks:167,RefWorks:801} The dimensions of the nanomagnet is 100 nm $\times$ 90 nm $\times$ 6 nm, which ensures that the nanomagnet has a single ferromagnetic domain.~\cite{RefWorks:402,RefWorks:133} With the chosen dimensions, the Stoner-Wohlfarth switching field $H_k$ becomes $\sim$$0.01\,M_s$. The values of stress ($\sigma_{max}$), strain ($\epsilon$), and asymmetric field ($H$) are 60.6 MPa, 433$\times$10$^{-6}$, and 0.01 T, respectively. 

Figure~\ref{fig:distribution_energy_300K} shows the distribution of energy dissipation at 300 K during the time interval $t_2 - t_1$ = 100 ns,  when stress is ramped up from zero to the maximum value, which makes the potential landscape monostable and erases the bit of stored information. This decreases the entropy of the system by $k\,ln(2)=0.6932\,k$. A concomitant amount of energy must be dissipated according to Landauer's principle. Stochastic violation of Landauer's bound 0.6932$\,kT$ due to thermal fluctuations is observed but the \emph{mean} energy dissipation does respect the Landauer's bound. This signifies the \emph{generalized} Landauer principle for \emph{small} systems, which are prone to thermal fluctuations. The mean energy dissipation during the time interval $t_3 - t_2$ (when magnetization traverses from $\theta=90^\circ$ towards $\theta=0^\circ$) is 0.07$\,kT$, which does not have any bound since there is no entropy loss in the system.

\section{CONCLUSIONS} 

Multiferroic straintronic devices turn out to be promising for energy-efficient computing in beyond Moore's law era. Since stress anisotropy in the magnetostrictive nanomagnets contribute a \emph{symmetric} term, it is necessary to have an \emph{asymmetric} component to maintain the direction of switching. Intrinsic magnetization dynamics can contribute to such \emph{asymmetric} term. With realistic parameters, the slight out-of-plane excursion of magnetization plays a crucial role in shaping the magnetization dynamics and therefore it is of paramount importance to consider complete three-dimensional potential landscape and solve full three-dimensional magnetization dynamics rather than \emph{assuming} magnetization is always confined to magnet's plane. Also intrinsic coupling between the polarization and magnetization can harness the \emph{asymmetric} term. These straintronic devices can be utilized not only for digital computing but also for analog signal processing. We have shown as well that the ultimate limit (so-called Landauer limit) of energy dissipation is achievable in a magnetostrictive nanomagnet, linking information and thermodynamics. These findings would hopefully stimulate experimental efforts to demonstrate that the magnetostrictive nanomagnets are suitable for exploring the thermodynamic limit of energy dissipation. The miniscule energy dissipation in these straintronic devices can be the basis of ultra-low-energy computing for our future information processing systems. This can open up as well some unprecedented applications that need to work with the energy harvested from the environment e.g., monitoring an epileptic patient's brain to notify an impending seizure by drawing energy solely from the patient's body.


\acknowledgments     
 
This work was supported in part by FAME, one of six centers of STARnet, a Semiconductor Research Corporation program sponsored by MARCO and DARPA.



\begin{thebibliography}{10}

\bibitem{roy13_spin}
K.~Roy, ``Ultra-low-energy straintronics using multiferroic composites,'' {\em
  SPIN}~{\bf 3}(2), p.~1330003, 2013.

\bibitem{roy11_news}
K.~Roy, S.~Bandyopadhyay, and J.~Atulasimha, ``Hybrid spintronics and
  straintronics: A magnetic technology for ultra low energy computing and
  signal processing,'' {\em Appl. Phys. Lett.}~{\bf 99}(6), p.~063108, 2011.
\newblock \\News: ``Switching up spin,'' \it{Nature} \bf{476}\normalfont, 375
  (Aug. 25, 2011), doi:10.1038/476375c.

\bibitem{RefWorks:558}
N.~A. Spaldin and M.~Fiebig, ``The renaissance of magnetoelectric
  multiferroics,'' {\em Science}~{\bf 309}(5733), pp.~391--392, 2005.

\bibitem{Refworks:164}
W.~Eerenstein, N.~D. Mathur, and J.~F. Scott, ``Multiferroic and
  magnetoelectric materials,'' {\em Nature}~{\bf 442}(7104), pp.~759--765,
  2006.

\bibitem{Refworks:165}
C.~W. Nan, M.~I. Bichurin, S.~Dong, D.~Viehland, and G.~Srinivasan,
  ``Multiferroic magnetoelectric composites: Historical perspective, status,
  and future directions,'' {\em J. Appl. Phys.}~{\bf 103}(3), p.~031101, 2008.

\bibitem{RefWorks:519}
N.~A. Pertsev, ``Giant magnetoelectric effect via strain-induced spin
  reorientation transitions in ferromagnetic films,'' {\em Phys. Rev. B}~{\bf
  78}(21), p.~212102, 2008.

\bibitem{roy11_6}
K.~Roy, S.~Bandyopadhyay, and J.~Atulasimha, ``Energy dissipation and switching
  delay in stress-induced switching of multiferroic nanomagnets in the presence
  of thermal fluctuations,'' {\em J. Appl. Phys.}~{\bf 112}(2), p.~023914,
  2012.

\bibitem{roy13_2}
K.~Roy, S.~Bandyopadhyay, and J.~Atulasimha, ``Binary switching in a
  `symmetric' potential landscape,'' {\em Sci. Rep.}~{\bf 3}(3038), p.~3038,
  2013.

\bibitem{roy11_2}
K.~Roy, S.~Bandyopadhyay, and J.~Atulasimha, ``Switching dynamics of a
  magnetostrictive single-domain nanomagnet subjected to stress,'' {\em Phys.
  Rev. B}~{\bf 83}(22), p.~224412, 2011.

\bibitem{roy13}
K.~Roy, ``Ultra-low-energy non-volatile straintronic computing using single
  multiferroic composites,'' {\em Appl. Phys. Lett.}~{\bf 103}(17), p.~173110,
  2013.

\bibitem{roy14}
K.~Roy, ``Critical analysis and remedy of switching failures in straintronic
  logic using bennett clocking in the presence of thermal fluctuations,'' {\em
  Appl. Phys. Lett.}~{\bf 104}(1), p.~013103, 2014.

\bibitem{RefWorks:559}
N.~Tiercelin, Y.~Dusch, A.~Klimov, S.~Giordano, V.~Preobrazhensky, and
  P.~Pernod, ``Room temperature magnetoelectric memory cell using
  stress-mediated magnetoelastic switching in nanostructured multilayers,''
  {\em Appl. Phys. Lett.}~{\bf 99}(19), p.~192507, 2011.

\bibitem{RefWorks:806}
M.~Liu, S.~Li, Z.~Zhou, S.~Beguhn, J.~Lou, F.~Xu, T.~J. Lu, and N.~X. Sun,
  ``Electrically induced enormous magnetic anisotropy in {Terfenol-D}/lead zinc
  niobate-lead titanate multiferroic heterostructures,'' {\em J. Appl.
  Phys.}~{\bf 112}(6), p.~063917, 2012.

\bibitem{RefWorks:609}
N.~Lei, T.~Devolder, G.~Agnus, P.~Aubert, L.~Daniel, J.~Kim, W.~Zhao,
  T.~Trypiniotis, R.~P. Cowburn, L.~Daniel, D.~Ravelosona, and P.~Lecoeur,
  ``Strain-controlled magnetic domain wall propagation in hybrid
  piezoelectric/ferromagnetic structures,'' {\em Nature Commun.}~{\bf 4}(1378),
  pp.~1378--1--1378--7, 2013.

\bibitem{RefWorks:790}
T.~Jin, L.~Hao, J.~Cao, M.~Liu, H.~Dang, Y.~Wang, D.~Wu, J.~Bai, and F.~Wei,
  ``Electric field control of anisotropy and magnetization switching in {CoFe}
  and {CoNi} thin films for magnetoelectric memory devices,'' {\em Appl. Phys.
  Express}~{\bf 7}(4), p.~043002, 2014.

\bibitem{RefWorks:820}
A.~Chopra, E.~Panda, Y.~Kim, M.~Arredondo, and D.~Hesse, ``Epitaxial
  ferroelectric {$Pb(Mg_{1/3}Nb_{2/3})O_3$-$PbTiO_3$} thin films on
  {$La_{0.7}Sr_{0.3}MnO_3$} bottom electrode,'' {\em J. Electroceram.} ,
  pp.~1--5, 2014.
\newblock \\DOI: 10.1007/s10832-014-9936-y.

\bibitem{razav01}
B.~Razavi, {\em Design of Analog {CMOS} Integrated Circuits}, McGraw-Hill Inc.,
  New York, NY, 2001.

\bibitem{unpub2x}
{K. Roy,} unpublished.

\bibitem{RefWorks:577}
M.~Julliere, ``Tunneling between ferromagnetic films,'' {\em Phys. Lett.
  A}~{\bf 54}(3), pp.~225--226, 1975.

\bibitem{RefWorks:33}
S.~S.~P. Parkin, C.~Kaiser, A.~Panchula, P.~M. Rice, B.~Hughes, M.~G. Samant,
  and S.~H. Yang, ``Giant tunnelling magnetoresistance at room temperature with
  {MgO} (100) tunnel barriers,'' {\em Nature Mater.}~{\bf 3}(12), pp.~862--867,
  2004.

\bibitem{RefWorks:581}
T.~Graf, S.~S.~P. Parkin, and C.~Felser, ``{Heusler Compounds--A Material Class
  With Exceptional Properties},'' {\em IEEE Trans. Magn.}~{\bf 47}(2),
  pp.~367--373, 2011.

\bibitem{RefWorks:162}
L.~Landau and E.~Lifshitz, ``On the theory of the dispersion of magnetic
  permeability in ferromagnetic bodies,'' {\em Phys. Z. Sowjet.}~{\bf 8}(153),
  pp.~101--114, 1935.

\bibitem{RefWorks:161}
T.~L. Gilbert, ``A phenomenological theory of damping in ferromagnetic
  materials,'' {\em IEEE Trans. Magn.}~{\bf 40}(6), pp.~3443--3449, 2004.

\bibitem{RefWorks:186}
W.~F. Brown, ``Thermal fluctuations of a single-domain particle,'' {\em Phys.
  Rev.}~{\bf 130}(5), pp.~1677--1686, 1963.

\bibitem{RefWorks:557}
E.~C. Stoner and E.~P. Wohlfarth, ``A mechanism of magnetic hysteresis in
  heterogeneous alloys,'' {\em Phil. Trans. Roy. Soc. A (London)}~{\bf 240},
  pp.~599--642, 1948.

\bibitem{RefWorks:157}
S.~Chikazumi, {\em {Physics of Magnetism}}, Wiley New York, 1964.

\bibitem{RefWorks:402}
M.~Beleggia, M.~D. Graef, Y.~T. Millev, D.~A. Goode, and G.~E. Rowlands,
  ``Demagnetization factors for elliptic cylinders,'' {\em J. Phys. D: Appl.
  Phys.}~{\bf 38}(18), pp.~3333--3342, 2005.

\bibitem{RefWorks:167}
T.~Brintlinger, S.~H. Lim, K.~H. Baloch, P.~Alexander, Y.~Qi, J.~Barry,
  J.~Melngailis, L.~Salamanca-Riba, I.~Takeuchi, and J.~Cumings, ``In situ
  observation of reversible nanomagnetic switching induced by electric
  fields,'' {\em Nano Lett.}~{\bf 10}(4), pp.~1219--1223, 2010.

\bibitem{RefWorks:801}
J.~Lou, D.~Reed, C.~Pettiford, M.~Liu, P.~Han, S.~Dong, and N.~X. Sun, ``Giant
  microwave tunability in {FeGaB}/lead magnesium niobate-lead titanate
  multiferroic composites,'' {\em Appl. Phys. Lett.}~{\bf 92}(26), p.~262502,
  2008.

\bibitem{RefWorks:821}
S.~M.~M. Quintero, C.~Martelli, A.~Braga, L.~C.~G. Valente, and C.~C. Kato,
  ``Magnetic field measurements based on terfenol coated photonic crystal
  fibers,'' {\em Sensors}~{\bf 11}(12), pp.~11103--11111, 2011.

\bibitem{roy14_2}
K.~Roy, ``Electric field-induced magnetization switching in interface-coupled
  multiferroic heterostructures: a highly-dense, non-volatile, and
  ultra-low-energy computing paradigm,'' {\em J. Phys. D: Appl. Phys.}~{\bf
  47}(25), p.~252002, 2014.

\bibitem{RefWorks:649}
M.~Fechner, P.~Zahn, S.~Ostanin, M.~Bibes, and I.~Mertig, ``Switching
  magnetization by 180$^\circ$ with an electric field,'' {\em Phys. Rev.
  Lett.}~{\bf 108}(19), p.~197206, 2012.

\bibitem{unpub}
{K. Roy,} unpublished.

\bibitem{RefWorks:144}
C.~H. Bennett, ``The thermodynamics of computation - a review,'' {\em Int. J.
  Theor. Phys.}~{\bf 21}(12), pp.~905--940, 1982.

\bibitem{RefWorks:148}
R.~Landauer, ``Irreversibility and heat generation in the computing process,''
  {\em IBM J. Res. Dev.}~{\bf 5}(3), pp.~183--191, 1961.

\bibitem{maxwell}
J.~C. Maxwell, {\em Theory of Heat}, Appleton, London, 1871.

\bibitem{RefWorks:615}
K.~Maruyama, F.~Nori, and V.~Vedral, ``The physics of {M}axwell's demon and
  information,'' {\em Rev. Mod. Phys.}~{\bf 81}(1), pp.~1--23, 2009.

\bibitem{RefWorks:619}
H.~Leff and A.~F. Rex, {\em Maxwell's Demon 2 Entropy, Classical and Quantum
  Information, Computing}, vol.~2, CRC Press, 2010.

\bibitem{RefWorks:642}
D.~Mandal and C.~Jarzynski, ``Work and information processing in a solvable
  model of maxwellӳ demon,'' {\em Proc. Nat. Acad. Sci.}~{\bf 109}(29),
  pp.~11641--11645, 2012.

\bibitem{RefWorks:614}
L.~Szilard, ``On the decrease of entropy in a thermodynamic system by the
  intervention of intelligent beings,'' {\em Z. Phys.}~{\bf 53}, pp.~840--856,
  1929.

\bibitem{RefWorks:627}
L.~Brillouin, ``Maxwell's demon cannot operate: Information and entropy. {I},''
  {\em J. of Appl. Phys.}~{\bf 22}(3), pp.~334--337, 1951.

\bibitem{RefWorks:528}
C.~Shannon, ``The mathematical theory of communication,'' {\em Bell Syst. Tech.
  J.}~{\bf 27}(7), pp.~379--423, 1948.

\bibitem{RefWorks:527}
R.~Landauer, ``Minimal energy requirements in communication,'' {\em
  Science}~{\bf 272}(5270), pp.~1914--1918, 1996.

\bibitem{RefWorks:645}
J.~V. Neumann, ``Theory of self-reproducing automata,'' {\em Univ. Illinois
  Press, Urbana, IL} , 1966.

\bibitem{RefWorks:613}
R.~Landauer, ``Dissipation and noise immunity in computation and
  communication,'' {\em Nature}~{\bf 335}(6193), pp.~779--784, 1988.

\bibitem{RefWorks:522}
C.~H. Bennett, ``Logical reversibility of computation,'' {\em IBM J. Res.
  Dev.}~{\bf 17}(6), pp.~525--532, 1973.

\bibitem{RefWorks:644}
C.~H. Bennett and R.~Landauer, ``The fundamental physical limits of
  computation,'' {\em Sci. Am.}~{\bf 253}(1), pp.~48--56, 1985.

\bibitem{RefWorks:629}
C.~Jarzynski, ``Nonequilibrium equality for free energy differences,'' {\em
  Phys. Rev. Lett.}~{\bf 78}(14), p.~2690, 1997.

\bibitem{RefWorks:637}
G.~E. Crooks, ``Entropy production fluctuation theorem and the nonequilibrium
  work relation for free energy differences,'' {\em Phys. Rev. E}~{\bf 60}(3),
  p.~2721, 1999.

\bibitem{RefWorks:626}
C.~Bustamante, J.~Liphardt, and F.~Ritort, ``The nonequilibrium thermodynamics
  of small systems,'' {\em Phys. Today}~{\bf 58}, p.~43, 2005.

\bibitem{RefWorks:623}
R.~Dillenschneider and E.~Lutz, ``Memory erasure in small systems,'' {\em Phys.
  Rev. Lett.}~{\bf 102}(21), p.~210601, 2009.

\bibitem{RefWorks:624}
B.~Lambson, D.~Carlton, and J.~Bokor, ``Exploring the thermodynamic limits of
  computation in integrated systems: Magnetic memory, nanomagnetic logic, and
  the {L}andauer limit,'' {\em Phys. Rev. Lett.}~{\bf 107}(1), p.~010604, 2011.

\bibitem{RefWorks:643}
T.~Sagawa, ``Thermodynamics of information processing in small systems,'' {\em
  Prog. Theor. Phys.}~{\bf 127}(1), pp.~1--56, 2012.

\bibitem{RefWorks:621}
G.~M. Wang, E.~M. Sevick, E.~Mittag, D.~J. Searles, and D.~J. Evans,
  ``Experimental demonstration of violations of the second law of
  thermodynamics for small systems and short time scales,'' {\em Phys. Rev.
  Lett.}~{\bf 89}(5), p.~050601, 2002.

\bibitem{RefWorks:622}
D.~M. Carberry, J.~C. Reid, G.~M. Wang, E.~M. Sevick, D.~J. Searles, and D.~J.
  Evans, ``Fluctuations and irreversibility: An experimental demonstration of a
  second-law-like theorem using a colloidal particle held in an optical trap,''
  {\em Phys. Rev. Lett.}~{\bf 92}(14), p.~140601, 2004.

\bibitem{RefWorks:620}
S.~Toyabe, T.~Sagawa, M.~Ueda, E.~Muneyuki, and M.~Sano, ``Experimental
  demonstration of information-to-energy conversion and validation of the
  generalized {J}arzynski equality,'' {\em Nat. Phys.}~{\bf 6}(12),
  pp.~988--992, 2010.

\bibitem{RefWorks:617}
A.~B\'{e}rut, A.~Arakelyan, A.~Petrosyan, S.~Ciliberto, R.~Dillenschneider, and
  E.~Lutz, ``Experimental verification of {L}andauer's principle linking
  information and thermodynamics,'' {\em Nature}~{\bf 483}(7388), pp.~187--189,
  2012.

\bibitem{RefWorks:490}
W.~F. Brown, ``The fundamental theorem of fine ferromagnetic particle theory,''
  {\em J. Appl. Phys.}~{\bf 39}, p.~993, 1968.

\bibitem{RefWorks:133}
R.~P. Cowburn, D.~K. Koltsov, A.~O. Adeyeye, M.~E. Welland, and D.~M. Tricker,
  ``Single-domain circular nanomagnets,'' {\em Phys. Rev. Lett.}~{\bf 83}(5),
  pp.~1042--1045, 1999.

\bibitem{RefWorks:426}
V.~Skumryev, S.~Stoyanov, Y.~Zhang, G.~Hadjipanayis, D.~Givord, and J.~Nogués,
  ``Beating the superparamagnetic limit with exchange bias,'' {\em Nature}~{\bf
  423}(6942), pp.~850--853, 2003.

\bibitem{kilby}
J.~S. Kilby, {\em Nobel Lecture in Physics}, The Nobel Foundation, Sweden,
  2000.

\bibitem{moore65}
G.~E. Moore, ``{Cramming More Components onto Integrated Circuits},'' {\em
  Proc. IEEE}~{\bf 86}, pp.~82--85, Jan. 1998.

\bibitem{RefWorks:211}
S.~Borkar, ``Design challenges of technology scaling,'' {\em IEEE Micro}~{\bf
  19}(4), pp.~23--29, 1999.

\end{thebibliography}

\end{document}